\documentclass[aip,jap,groupedaddress,reprint]{revtex4-1}
\usepackage{graphicx}
\usepackage{ulem}
\usepackage{amssymb}
\usepackage{color}

\begin{document}


\title{Calculating field emission currents in 
nanodiodes - a multi-group formalism with space charge and exchange-correlation effects}
 

\author{Debabrata Biswas}
\author{Raghwendra Kumar} 

\affiliation{Theoretical Physics Division,
Bhabha Atomic Research Centre,
Mumbai 400 085, INDIA}


\begin{abstract}
Inclusion of electron-electron interaction is essential
in nano-diodes to understand the underlying physical phenomenon
and tailor devices accordingly. However,
both space charge and exchange-correlation interaction
involve electrons 
at different energies and hence a self-consistent
multi-energy-group solution of the Schr\"{o}dinger-Poisson system
is required. It is shown here that the existence of 
a limiting density-dependent potential at low
applied voltages allows calculation of the field
emission current. Despite additional interactions, a Fowler-Nordheim 
behaviour is observed.
It is also found that the exchange-correlation potential dominates
at these voltages in nanogaps and possibly leads to a  higher
turn-on voltage.
\end{abstract}






\maketitle

\newcommand{\be}{\begin{equation}}
\newcommand{\ee}{\end{equation}}
\newcommand{\bea}{\begin{eqnarray}}
\newcommand{\eea}{\end{eqnarray}}
\newcommand{\Tbar}{{\bar{T}}}
\newcommand{\En}{{\cal E}}
\newcommand{\Lop}{{\cal L}}
\newcommand{\DB}[1]{\marginpar{\footnotesize DB: #1}}
\newcommand{\q}{\vec{q}}
\newcommand{\kt}{\tilde{k}}
\newcommand{\Lopn}{\tilde{\Lop}}
\newcommand{\noi}{\noindent}
\newcommand{\ovn}{\bar{n}}
\newcommand{\ovx}{\bar{x}}
\newcommand{\ovE}{\bar{E}}
\newcommand{\ovV}{\bar{V}}
\newcommand{\ovU}{\bar{U}}
\newcommand{\ovJ}{\bar{J}}
\newcommand{\calE}{{\cal E}}
\newcommand{\ovphi}{\bar{\phi}}



\section{Introduction}
\label{sec:intro}
Nanotechnology has brought to the fore exotic materials
such as carbon nanotubes and silicon nanowires that
are being researched for a wide range of applications.
Of particular interest is electron field-emission from
nano materials that has potential to be used in
vacuum electronics \cite{flat_panel}, lithography and microwave power
amplifiers \cite{microwave}. Factors such as high efficiency, high current, 
low turn on voltages and fast turn-on times are being
investigated for integration into existing devices.
The necessity of an appropriate theoretical framework to deal with
electron emission at the nanoscale is thus very important.

It is well known that field emission is an inherently 
quantum-mechanical phenomenon \cite{FN} governed by the 
Fowler-Nordheim (FN) law:

\be
J =  \frac{A}{\phi} (\En^2) \exp(-B \phi^{3/2}/\En) \label{eq:FN}
\ee

\noi
where $A$ and $B$ are constants, $\phi$ is the work function
and $\En = V_g/D$ is the applied electric field where
$V_g$ is the applied voltage and $D$ the spacing between electrodes. 
A signature straight line 
fit in a $\ln(J/V_g^2)$ vs $1/V_g$ plot demonstrates that
the simplified tunneling through a triangular-barrier 
model captures the essential physics,
even though the calculated currents
differ significantly from the observed ones.

Modifications to the law generally concentrate on the
enhancement of the applied field \cite{forbes,Wang} due to surface 
morphology and the image potential \cite{Nordheim,jensen99,
rokhlenko11,jensen2003,rokhlenko2010}. 
Most experimental studies characterize emitters using the 
effective work function and the enhancement factor $\beta$ from an FN-plot
with $\En$ replaced by $\beta \En$ in Eq.~(\ref{eq:FN}) above.
However the fitted value of $\beta$ and field enhancement 
calculations using observed surface morphologies seldom
match \cite{wisniewski}, leaving much scope for including 
physical phenomenon hitherto kept aside to maintain simplicity.

At the nanoscale especially, space charge and 
exchange-correlation potentials
are important and can significantly
alter the turn-on voltage. While these effects
are small in a ``standard'' measure-theoretic sense compared
to the work-function, the image potential or the enhanced electrostatic
potential, these slight changes in the barrier height or width
can alter the transmission coefficient (and hence the emitted current)
by orders of magnitude. These interactions however 
involve electrons at different energies so that the 
electron emission at the nanoscale is essentially a
multi-energy group problem.

In the following, we shall formulate the field emission problem
by including the space-charge and exchange-correlation potentials,
discuss the challenges of multi-group emission and
show that an alternate approach allows
us to calculate the emission current at least at low applied voltages. 


\section{Formalism}
\label{sec:formalism}

Field emission calculations that take into account electron-electron interaction 
generally involve a two-step process. In the absence of these interactions, the potential
through which electrons tunnel into vacuum is independent of the 
current flowing through the gap. Thus, the density of electron 
states in the cathode and the transmission coefficient, both of which are known in-principle, 
determine the field emission current.
When electron-electron interaction is turned on, the transmission coefficient
depends on the current itself since the tunneling potential now depends on $J$. 
To deal with such a situation \cite{koh2006}, 
one may look for a solution to the equation

\be
J = \frac{e}{2\pi\hbar} \int T(E,J) f(E) dE  \label{eq:basic_emission}
\ee

\noi
where $e$ is the electron charge, $f(E)$ is the electron ``supply function''
within the emitter and $T(E,J)$ is the transmission coefficient which in 
turn depends on the emission current density $J$. This requires
(i) solving the Schr\"{o}dinger-Poisson system to determine the self-consistent
potential $V_{eff}$ assuming that the wavefunction carries a given current density $J$ 
and then (ii) using this potential to determine the transmission coefficient
and thus the emission current density by computing the 
integral in Eq.~(\ref{eq:basic_emission}). This process then can be repeated
till a solution is obtained.

As a first approximation, a single Schr\"{o}dinger equation carrying a current $J_k$ 
at an energy $E_k$ may be coupled to the Poisson equation in order to determine $V_{eff}$,
effectively implying that the calculation of the effective potential
is performed assuming that all the electrons are at a single energy.
More correctly, since the tunneling electrons are distributed across a range of 
energies, a multi-energy group approach needs to be adopted as we shall elaborate
later. Even without this complication however,
the approach outlined above is not guaranteed to lead to a 
solution to Eq.~(\ref{eq:basic_emission}).

To understand this, note that a self-consistent solution of the 
1-Schr\"{o}dinger-Poisson system
does not exist at any arbitrary value of $J$. In other words, there is a limiting
mechanism that allows a self-consistent solution only upto a maximum current-density
$J_k^{max}$ at an energy $E_k$ \cite{ang2003,epj2012,epl2013}. Note that at $J_k^{max}$, the tunneling
electrons see a broader and higher effective potential barrier compared 
to values of $J_k < J_k^{max}$. Despite this,  
the integral in  Eq.~(\ref{eq:basic_emission}) evaluated using the
effective potential at $J_k^{max}$, 
may lead to a value much larger than $J_k^{max}$ depending on the
supply function and the transmission coefficient.
Thus, a solution to Eq.~(\ref{eq:basic_emission}) may not exist especially 
in nano-diodes.
Note that the problem is not linked to the use of a mono-energetic electron density in the
Schr\"{o}dinger-Poisson system. Rather, while a multi-energy group approach in solving
the Schr\"{o}dinger-Poisson system is necessary, it is generally the case that
in nano-diodes, the exchange-correlation interaction lowers the maximum current
density that can be supported in the diode resulting in a higher transmission 
coefficient \cite{contrast}.
The method for evaluating the
field emission current at the nano level thus needs re-examination since
a solution to Eq~(\ref{eq:basic_emission}) may not exist.

Our formulation of the field emission problem for nano-diodes in the presence of electron-electron
interaction is as follows: given an applied voltage $V_g$ across the diode, the 
{\it maximum} current that can tunnel through and propagate across the gap in a 
self-consistent manner is the field-emission current. We shall use this definition
in a multi-energy group formalism to determine the cold field emission current from metals.

In metals, electrons are distributed from the bottom of the 
conduction band to the Fermi level. It is thus necessary 
to divide the energy of the emitted electrons 
in the gap  (vacuum) between 
the two metal electrodes, into $N$ groups \cite{Ninfinite} 
and solve $N$ Schr\"{o}dinger equations, one for
each energy group, coupled to each other through the Poisson equation
and the exchange-correlation potential:

\bea
-\frac{\hbar^2}{2m}\frac{d^2 \psi_k}{dx^2} + V_{eff} \psi_k & = &  E_k \psi_k \label{eq:Schrod} \\
\frac{d^2 V}{dx^2} = \frac{e}{\epsilon_0} n(x) & = & \frac{e}{\epsilon_0} \sum_{k=1}^N |\psi_k(x)|^2. \label{eq:Poi}
\eea

\noi
Here, $V_{eff} = \phi + V_{im} -eV + V_{xc} \times E_H$, where $\phi$ is the work-function,
$E_H = e^2/(4\pi\epsilon_0 a_0)$ is the 
Hartree energy, $a_0$ the Bohr radius and $V_{im} = -e^2/16 \pi \epsilon_0 (x+x_0)$ 
is the image-charge potential with $x_0$ such that $\phi + V_{im} = 0$ at $x = 0$ \cite{x0,kiejna}. 
The exchange correlation potential \cite{KS,PZ}, $V_{xc}$
in the local density approximation takes the form 
$V_{xc} = \epsilon_{xc} - (r_s/3) d\epsilon_{xc}/dr_s$, where
$r_s = [3/(4\pi n(x))]^{1/3}$ is 
the Wigner-Seitz radius, $n(x)$ the electron 
number density,
$\epsilon_{xc} = \epsilon_{x} + \epsilon_{c}$ and 
$\epsilon_{x}  =  -(3/4) (\frac{3}{2\pi})^{2/3}\frac{1}{r_s}$, 
$\epsilon_{c}  =  \gamma/(1 + \beta_1 \sqrt{r_s} + \beta_2 r_s)$
with $\gamma = -0.1423$, $\beta_1 = 1.0529$
and $\beta_2 = 0.3334$. The parametrized form for the 
correlation energy density, $\epsilon_{c}$
is due to Perdew and Zunger \cite{PZ}.

Note that $V_{eff}$ refers to the self-consistent potential
in the gap region between the two electrodes. 
Thus, $V_{xc}$ is the exchange-correlation
potential  due to the electrons present in the gap 
while the image potential, $V_{im}$, is the
linear response of the cathode \cite{lang_kohn_73,kiejna,neglected_anode} to the  
`external' charges (steady state electrons) present in the gap.
The `gap' electrons in the vaccum region are thus modelled
fully by the Kohn-Sham density functional theory and 
their only interaction with the cathode is due to the
image potential \cite{model_all}.

It is convenient to write the wavefunctions 
\be
\psi_k = \sqrt{n_0}~ q_k(x) \exp(i \theta_k(x))
\ee

\noi
in terms 
of a real amplitude $q_k(x)$ and phase $\theta_k(x)$
and deal with equations for the amplitude $q_k$. Moreover, since we 
shall be dealing with scaled potentials and distance, it is  necessary to 
frame the Schr\"{o}dinger-Poisson system in dimensionless form \cite{ang2003}
using the characteristic density 
$n_0 = 2\epsilon_0 V_g/3eD^2$, the applied voltage $V_g$, the 
electron de Broglie wavelength $\lambda_0 =  \sqrt{\hbar^2/2meV_g}$, 
the gap distance $D$ and the Child-Langmuir current density
$J_{CL}$ \cite{CL}.
In terms of the
dimensionless normalized variables
$\ovx = x/D$, $\ovV = V/V_g$, $\lambda = D/\lambda_0$, 
$\epsilon_k = E_k/eV_g$, $\ovJ = J/J_{CL}$, $\ovV_{im} = V_{im}/(eV_g)$,
$\ovV_{xc} = V_{xc}\times E_H/(eV_g)$ and
$\ovphi = \phi/(eV_g)$,
the $N$-Schr\"{o}dinger and Poisson equations can be expressed respectively 
as

\bea
\frac{d^2q_k}{d\ovx^2}&  = &  - \lambda^2 [\epsilon_k + \ovV  - \ovV_{xc} - \ovphi - \ovV_{im} - 
\frac{4}{9} \frac{\ovJ_k^2}{q_k^4} ]q_k  \label{eq:Sch1} \\
\frac{d^2\ovV}{d\ovx^2} &  = &  \frac{2}{3} \sum_k q_k^2 \label{eq:Poi1}
\eea

\noi
where $k = 0,1,\ldots,N-1$.

Eqns. (\ref{eq:Sch1}) and (\ref{eq:Poi1}) are
complemented with appropriate boundary conditions for
the potential $\ovV$ \cite{comment} and the amplitude $q_k$ \cite{epl2013}.
Note that each of the wavefunctions $\psi_k$ is assumed to carry 
a current density 

\be
J_k = \frac{e\hbar}{2m i} (\psi_k \frac{\partial\psi_k^*}{\partial x} - 
\psi_k^* \frac{\partial\psi_k}{\partial x})
\ee

\noi
so that the equation for $q_k$ is independent of $\theta_k$.

A set of currents $\{J_k\}$ for which Eqns. (\ref{eq:Sch1}) and (\ref{eq:Poi1})
can be solved is a candidate for the field emission current 
provided the emitter is able to emit 
the respective current densities $J_k$ from each of the energy 
segments 

\be
J_k  \leq \frac{e}{2\pi\hbar} \int_{E_k}^{E_k+\Delta} T(E) f(E) dE~~~~ 
\forall~ k \label{eq:condition}
\ee

\noi
where the supply function $f(E) = (m k_B T/\pi\hbar^2)\ln[1 + \exp((E_f - E)/k_BT)]$  
\cite{FN,jensen2003}. Here, 
the total integration domain $[0,E_F]$ is divided into $N$ equal
energy segments of size $\Delta$. A solution set $\{J_k\}$ yielding a valid
$V_{eff}$ normally satisfies Eqns.~(\ref{eq:condition}) in the nano regime. If the opposite
holds (the inequality in Eq.~(\ref{eq:condition}) is not satisfied), the
current will be ``emission limited''. 

\begin{figure*}[tb]
\vskip 0.5 in
\hspace{-32.0em}
\includegraphics[width=6.75cm, angle=0]{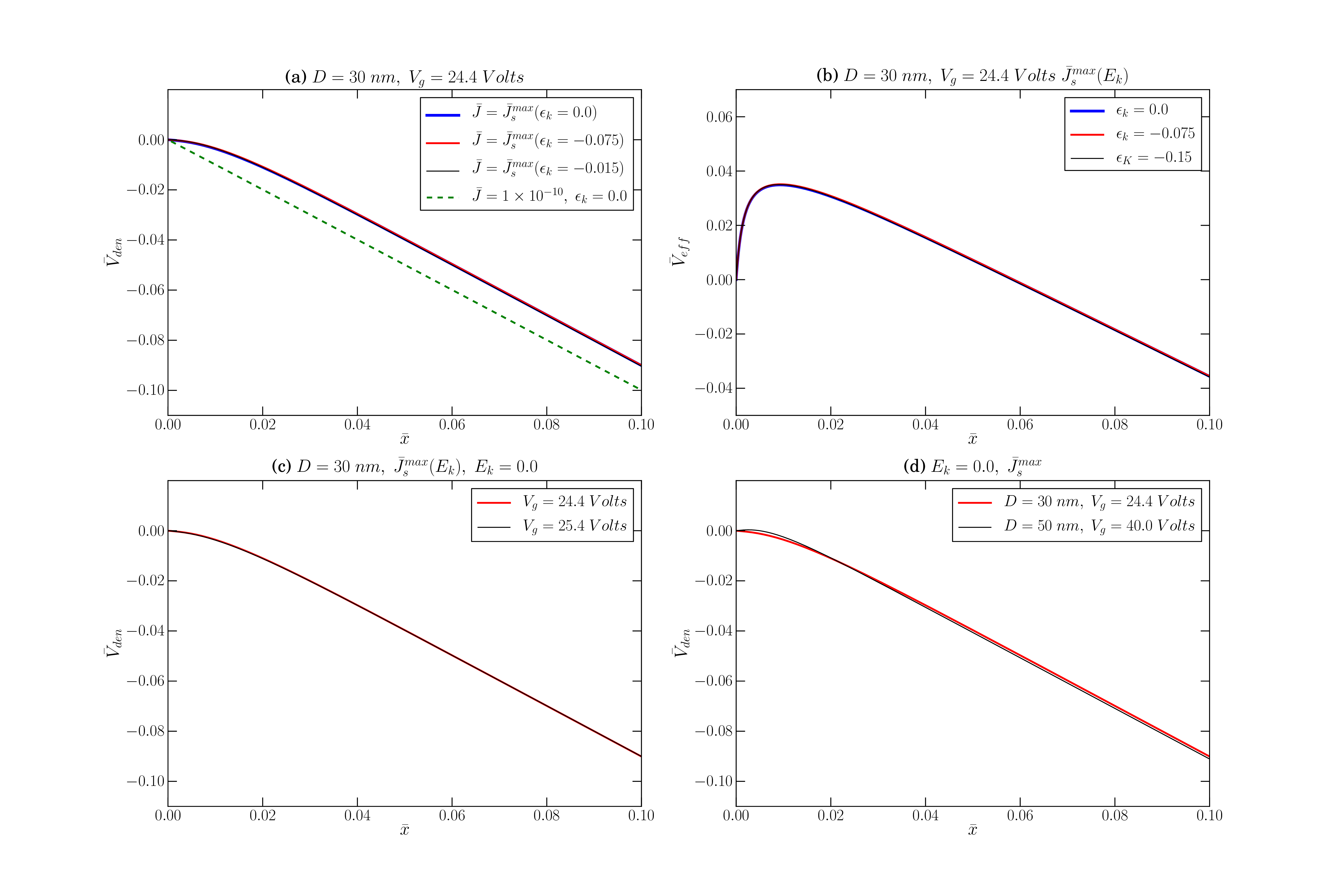}
\caption{The scaled density-dependent potential is shown
as a function of the scaled distance $\ovx$ in (a),(c) and (d). 
In (a), the dashed line corresponds
to $\ovJ << \ovJ_s^{max}(E_k=0)$ while 
$\ovJ_s^{max}(E_k=0) = 1.645\times 10^{-6}$, 
$J_s^{max}(E_k=-0.075)= 4.6\times10^{-7}$ and 
$J_s^{max}(E_k=-0.15)= 1.3\times10^{-7}$. In (b), the effective potential $V_{eff}$
at $D= 30$nm and $V_g = 24.4$V is compared for three different injection energies.
In (c), $\ovV_{den}$ is compared for two different voltages at the same $D$
while in (d) even the gap spacing is different.
Note that $\ovx$ varies from 0 to 1 and only a segment is shown to magnify the
differences. The energies $E_k$ are in units of $eV$.
}
\label{fig:1}
\end{figure*}

In keeping with our formulation of the field emission problem, 
we are interested here in the maximal set of current densities 
$\{J_k^{max}\}$ for which a solution of
Eqns. (\ref{eq:Sch1}) and (\ref{eq:Poi1}) exists and $J_k^{max}$ satisfies
the emission inequalities.
Determining this maximal set $\{J_k^{max}\}$ by solving simultaneously 
$N$ Schr\"{o}dinger equations and the Poisson equation, and checking for
its convergence is however a difficult task. We shall therefore 
take a different approach for determining the maximal set and hence the
field emission current.

\section{The Existence of a limiting potential}

The maximal set defines a limit beyond which steady state 
electron flow ceases to exist. In other words, the time-independent
Schr\"{o}dinger-Poisson system no longer yields a solution beyond
a limiting value for the current densities.
Importantly, the limiting mechanism is expected to be universal in the low voltage
``quantum regime'' as well at the high voltage ``classical regime''  
\cite{reflection,epl2013,garrido}.

In the low voltage quantum regime, such a universality can reflect in the 
variable or density-dependent part of the effective potential 
comprising of the space charge and exchange-correlation
potentials. To make a comparison, the scaled density-dependent 
voltage $\ovV_{den} = (V_{xc} \times E_H /e  - V)/V_g$ as a function of 
the scaled distance $\ovx$ needs to be studied.

To test the universality hypothesis, consider emission at a single 
energy $\epsilon_{k}$ (i.e. $N=1$). The maximum or limiting scaled current density at this
energy, $\ovJ_s^{max}(\epsilon_{k})$ (the subscript $s$ denoting single energy group), 
can be determined by solving a single Schr\"{o}dinger equation coupled to the
Poisson equation self-consistently. If a universal limiting mechanism hypothesis holds, 
a plot of $\ovV_{den}$ vs $\ovx$ should
be independent of $\epsilon_{k}$ at $\ovJ_s^{max}(\epsilon_{k})$. This is true as shown in 
Fig.~(\ref{fig:1}a) for $D = 30$nm, $V_g = 24.4$V for $E_{k} = 0$, 
$-0.075$ and $-0.15$. The potentials are indeed identical.  
To check that the space charge and exchange-correlation potentials
are not inconsequential at these low currents and charge densities, 
the scaled voltage  $\ovV_{den}$ is also compared
at these parameters for $\ovJ = 10^{-10}  << \ovJ_s^{max}$ where $J_{CL}$ is the 
Child-Langmuir current density \cite{CL}. At this value of $\ovJ$, the electron
density is indeed negligible and  $\ovV_{den}$  expectedly does not
bear the signature of the limiting potential.

The universality of this limiting potential is further tested by comparing
$\ovV_{den}$ at $V_g = 25.4$V and $V_g = 24.4$V for  $D = 30$nm at their respective
limiting current densities  $\ovJ_s^{max}(E_k=0;V_g)$. The change in voltage 
does not affect the scaled limiting potential as seen in Fig.~(\ref{fig:1}c).

Finally, in Fig.~(\ref{fig:1}d) a comparison of two different systems is
presented, one at $D = 30$nm and the other at $D=50$nm for $V_g = 24.4$V
and $V_g = 40$V respectively. In both cases, the scaled potential $\ovV_{den}$
is obtained for limiting current density  $\ovJ_s^{max}(0)$. Clearly, the 
limiting potentials are very close and may be considered identical.

For a given gap spacing and applied voltage, the scaled effective potential 
$\ovV_{eff} = V_{eff}/(eV_g)$
has in addition the scaled work-function $\ovphi$, image potential 
$\ovV_{im}$ and terms such as due to field enhancement 
(see section \ref{sec:results}), all of which
are identical at a given gap spacing $D$ and gap voltage $V_g$ at all energies.
Thus $\ovV_{eff}$ should also reflect the universality observed 
in the density-dependent potential. This is shown in Fig.~1b.

\section{The maximal Set and the limiting field emission current}
\label{sec:maximal}

There is thus ample numerical evidence to state that a limiting potential exists
which is independent of the electron energy when injection 
takes place at a single energy $\epsilon_k$ with a current density $\ovJ_s^{max}(\epsilon_k)$. 
Recall that beyond this maximal current, no solution exists for the
1-Schrodinger-Poisson system. The signature of the breakdown is the
limiting potential.

It is therefore reasonable to put forward the hypothesis 
that even in multi-group emission, for the maximal set $\{\ovJ_k^{max}\}$,
the potential $\ovV_{den}$, assumes the universal 1-group limiting form 
observed in Fig.~\ref{fig:1}.
Note that in each of the $N$ equations in  Eq.~(\ref{eq:Sch1}), 
the right hand side is identical to a 1-group limiting equation at energy $\epsilon_k$ 
if $\ovJ_k^{max} = w_k \ovJ_s^{max}(\epsilon_k)$
and $q_k(\ovx) = \sqrt{w_k}~ q_s^{max}(\ovx;\epsilon_k)$ with $w_k < 1$. Here $q_s^{max}(\ovx;\epsilon_k)$
is the 1-group amplitude at energy $\epsilon_k$ and current density $\ovJ_s^{max}(\epsilon_k)$.
With such a scaling, each of the $N$ Schrodinger equations assumes the respective
1-group limiting form at which steady state transmission ceases to exist.

\begin{figure}[tbh]
\vskip 0.4 in
\hspace{-15.0em}
\includegraphics[width=3.25cm, angle=0]{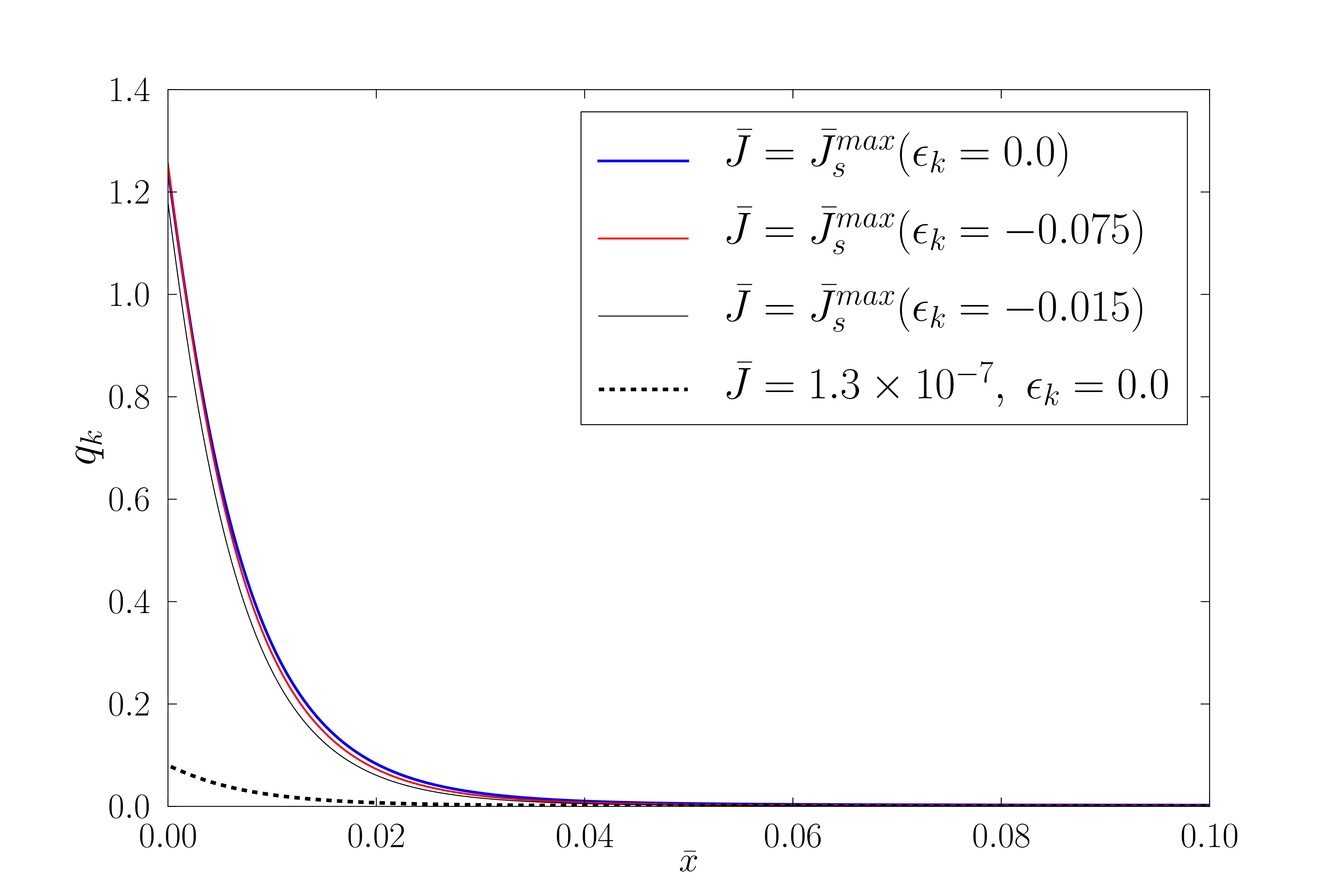}
\caption{The amplitude $q_k$ at $D= 30$nm and $V_g = 24.4$V for different
 currents and injection energies. Note that $q_k$ at $J_s^{max}(E_k = -0.15) = 1.30\times 10^{-7}$
attains the limiting form while for the same current at a higher energy (dashed line), $q_k$
is much smaller. The energies $E_k$ are in units of $eV$.
 }
\label{fig:ampl}
\end{figure}

Along with the $N$ Schr\"{o}dinger equations, the Poisson equation too should
assume the 1-group limiting form. That such a form exists is borne 
out by plotting the 1-group amplitudes $q_s(\ovx;\epsilon_k)$ at $\ovJ_s^{max}(\epsilon_k)$ 
for different injection
energies. At the limiting current, the amplitudes $q_s$ (and hence the 
charge density) assume a limiting form, $q_s^{max}$ irrespective of the injection energy
as can be seen in Fig.~\ref{fig:ampl}. In contrast, when $\ovJ_s(\epsilon_k) < \ovJ_s^{max}(\epsilon_k)$, 
the amplitude is far away from the limiting form as shown in Fig.~\ref{fig:ampl}.

In the multi-group case then, the Poisson equation can be written as 

\be
\frac{d^2\ovV}{d\ovx^2} = \frac{2}{3} \sum_k q_k^2(\ovx) = \frac{2}{3} 
\sum_k w_k~(q_s^{max}(\ovx;\epsilon_k))^2
\ee

\noi
when the  $N$ Schr\"{o}dinger equations assume the 1-group limiting form.
However, since $q_s^{max}(\ovx;\epsilon_k) = q_s^{max}(\ovx) $ is independent of $\epsilon_k$, 
we have 

\be
\frac{d^2\ovV}{d\ovx^2} =  \frac{2}{3} \sum_k w_k~(q_s^{max}(\ovx))^2 =  
\frac{2}{3}(q_s^{max}(\ovx))^2 \sum_k w_k .
\ee

\noi
It then follows on demanding that the limiting multi-group electron density be identical
to $(q_s^{max}(\ovx))^2$, that $\sum_k w_k~=~1$.

Thus, with $J_k = w_k J_s^{max}(\epsilon_k)$ and $q_k(\ovx) = \sqrt{w_k} q_s^{max}(\ovx)$, each of 
the $N$ Schr\"{o}dinger equations together with the Poisson equation assume the limiting
form where steady state transmission ceases to exist provided $\sum_k w_k~=~1$. 
Thus $\{ w_k J_s^{max}(\epsilon_k) \}$ is
the maximal set and the field emission
current is 

\be
J = \sum_k w_k J_s^{max}(\epsilon_k). \label{eq:Jweighted}
\ee

\noi
It now remains to determine the weights $w_k$.

From a pure
emission point of view, the  
current that can tunnel through a given potential, $V_{eff}$, in an 
energy segment $\Delta$ at $E_k$ is

\be
(e/2\pi\hbar) \int_{E_k}^{E_k+\Delta} T(E) f(E) dE 
\simeq (e/2\pi\hbar) T(E_k)f(E_k) \Delta
\ee

\begin{figure}[tbh]
\vskip 0.4 in
\hspace{-15.0em}
\includegraphics[width=3.25cm, angle=0]{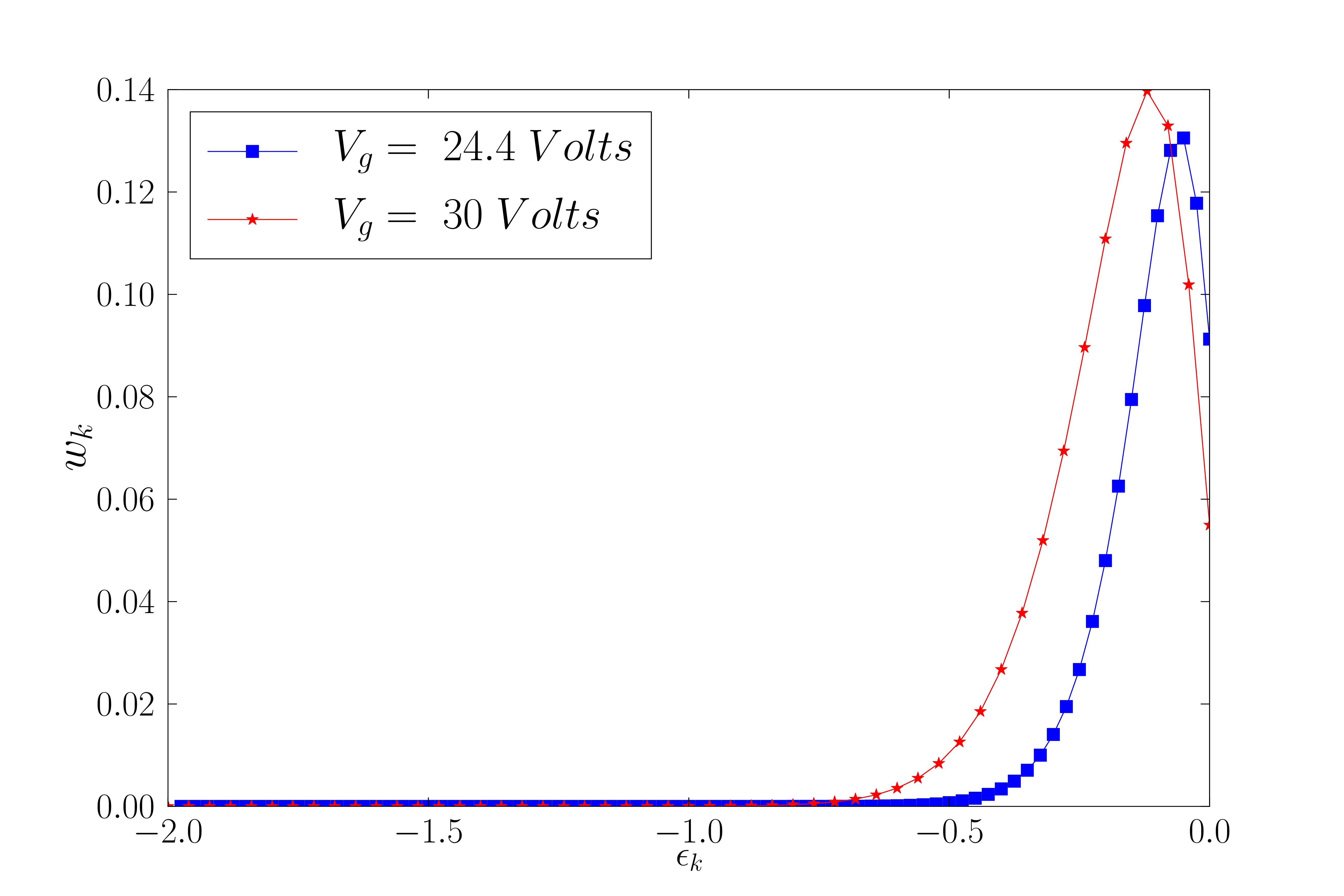}
\caption{The weights $w_k$ for different injection energies at two
different voltages. Here $D = 30$nm. The energies $E_k$ are in units of $eV$. 
 }
\label{fig:weights}
\end{figure}

\noi
if $N$ is large enough.
The weight of each segment is thus proportional to $T(E_k)f(E_k)$.
When the potential in question is the limiting effective potential, 
the weight of each energy segment contributing
to the charge density should again be proportional to  $T(E_k)f(E_k)$
where $T(E_k)$ is computed using the limiting potential.
Thus the normalized weights are
$w_k = T(E_k)f(E_k)/(\sum_k T(E_k)f(E_k))$ so that $\sum_k w_k = 1$. 
Since the limiting potential is known, the weights can be calculated
and the field emission current determined. Typical plots
of the weight at $D = 30$nm for two different applied voltages
are shown in Fig.~\ref{fig:weights}. 

Note that at higher applied voltages for a given nanogap (fixed $D$), the
limiting mechanism shifts and is increasingly dominated 
by the Poisson equation \cite{negligible}.
Arguments presented for the maximal set no longer hold in this
domain. 

\section{Numerical Results}
\label{sec:results}

We shall now use the prescription outlined in the earlier sections to 
determine the field emission current. 
The emitter used in this paper is Tungsten having $\phi = 4.55$eV
and $E_F = 10.46$ eV. Though the model used in this paper is 1-dimensional,
we introduce field enhancement artificially to mimic a realistic 
situation, using a 
scaled applied potential based on the floating sphere model \cite{Wang,choice}
by choosing the direction (angle $\theta = 0$) along the gap. 
Thus, when no charge is present, the 
scaled applied potential is taken to be

\bea
&~& \ovV_{enh}(x) = \frac{1}{4\pi\epsilon_0 V_g} [-\frac{{\cal D}}{(x+\rho)^2}
+ \frac{{\cal C}}{(x+2h+\rho)} \nonumber \\ 
& - & \frac{{\cal C}}{(x+\rho)} - \frac{{\cal D}}{(x+2h+\rho)^2)}] + \frac{E_a}{V_g} (x+h+\rho) \nonumber
\eea 

\begin{figure}[tbh]
\vskip 0.4 in
\hspace{-16.0em}
\includegraphics[width=3.60cm, angle=0]{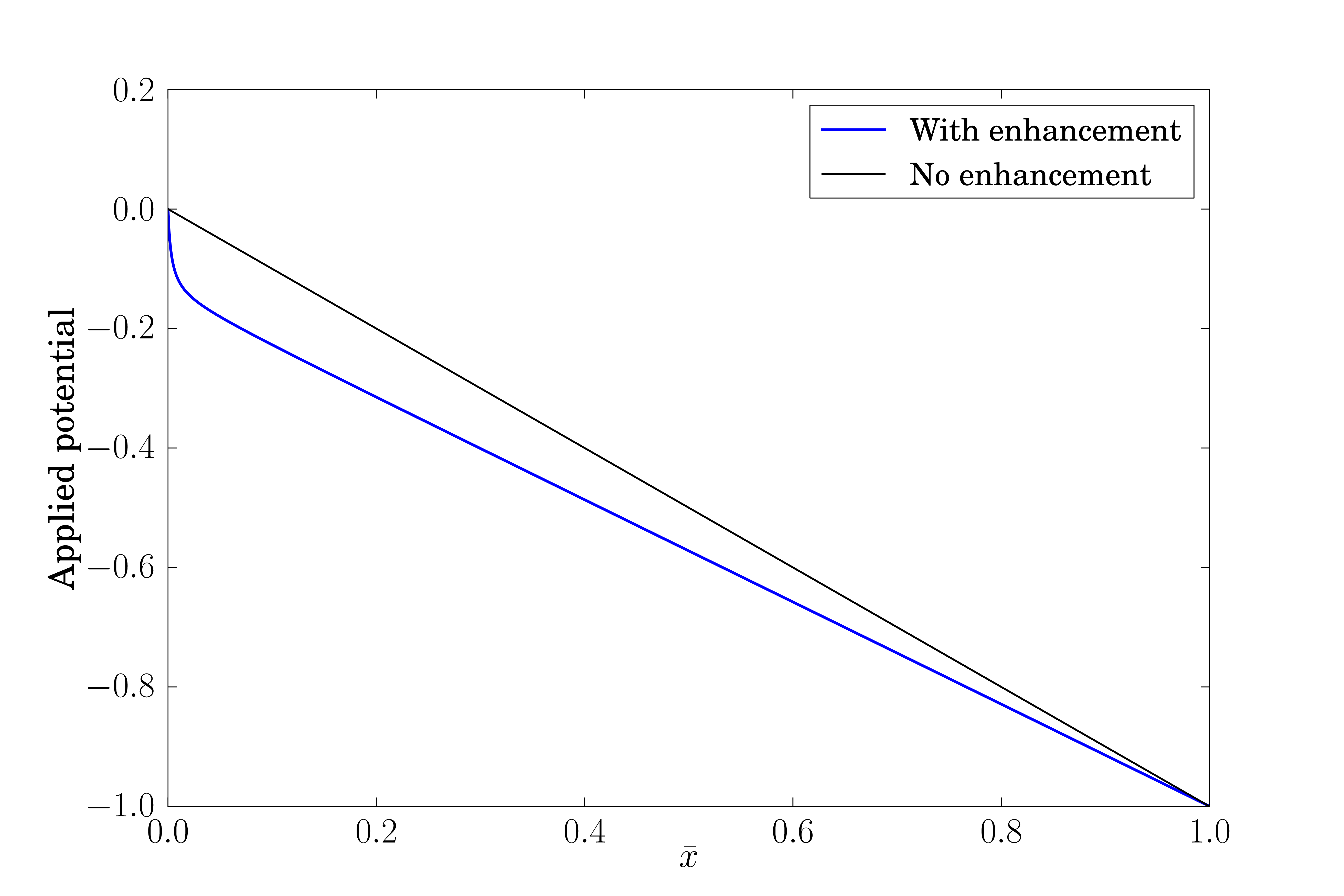}
\caption{A comparison of the scaled applied potential with (lower curve)
and without (straight line) enhancement.
}
\label{fig:enh}
\end{figure}

\noi
where $\rho$ is the radius of the sphere, $h$ is the distance
of the center of the sphere from the cathode, $E_a = V_g/(D + h + \rho)$,
${\cal C} = (4 \pi \epsilon_0) E_a h \rho (1 + \rho/2h)$
and ${\cal D} = (4 \pi \epsilon_0) E_a \rho^3$. The values
of $\rho$ and $h$ have been chosen to be $0.1$nm and $5$nm respectively.
A comparison of the scaled applied potential with and without enhancement
is shown in Fig.~\ref{fig:enh}.

\begin{figure}[tbh]
\vskip 0.4 in
\hspace{-16.0em}
\includegraphics[width=3.60cm, angle=0]{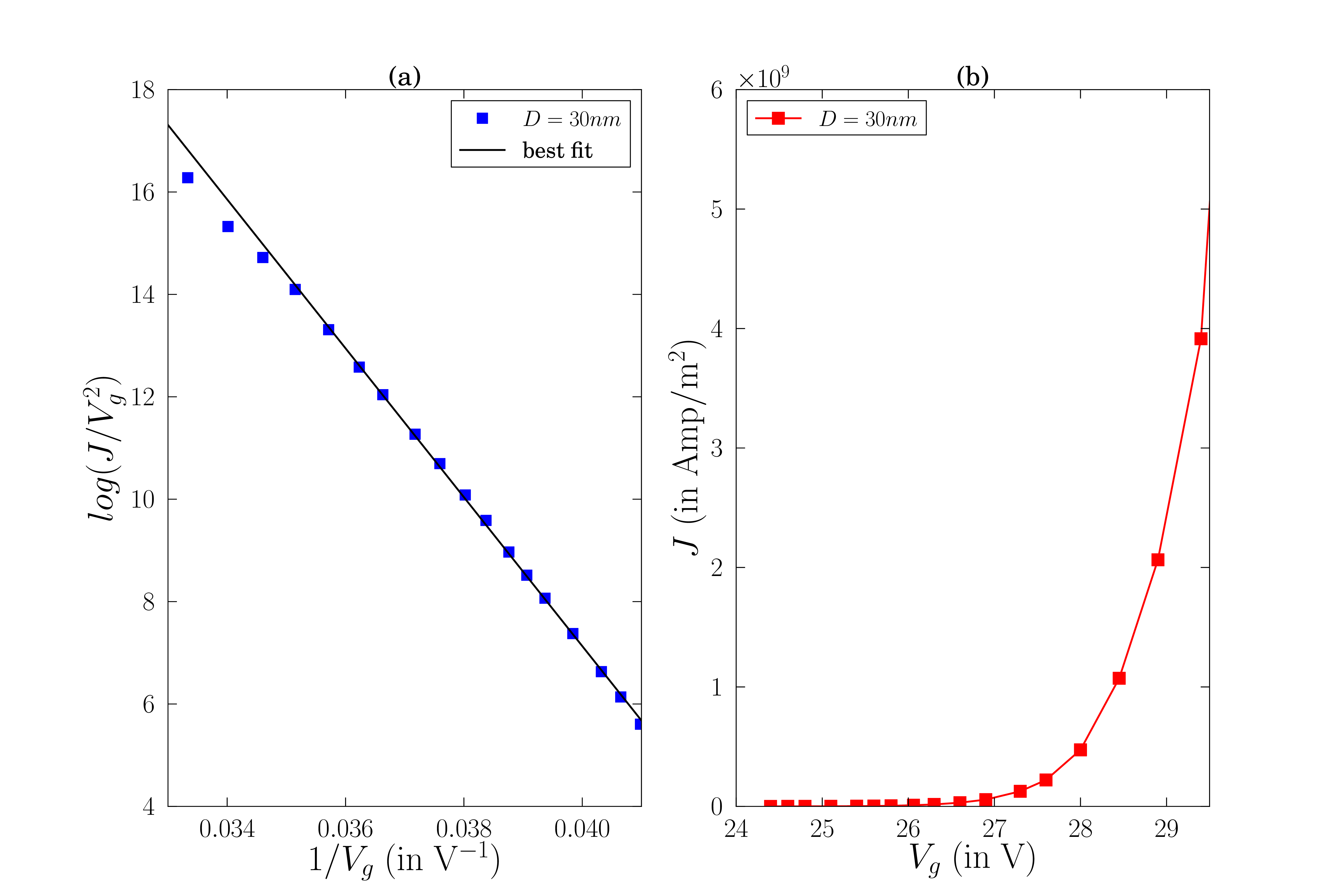}
\caption{The field emission current density evaluated 
using Eq.~(\ref{eq:Jweighted}) for $D = 30$nm. In (a) an FN plot
is shown along with the best fit $f(1/V_g)= -1453.97*(1/V_g)+65.29$.
while (b) is a normal current density vs applied voltage plot.
}
\label{fig:J}
\end{figure}

The limiting field-emission current density $J$ calculated
using Eq.~(\ref{eq:Jweighted}) is shown in Fig.~\ref{fig:J}. 
Clearly, at lower voltages, the emission
current follows an FN-curve as seen in experiments
despite the fact that the current density shown here
corresponds to the universal limiting potential.
At higher voltages, the current density moves away from
the FN-curve, coincident with the observation that 
$\ovV_{den}$ moves away from the limiting
potential thereby signalling a shift towards a limiting
mechanism dominated increasingly by the Poisson equation.
The current density should thus move towards the classical
Child-Langmuir law \cite{CL,pop3} at high applied voltages.

The effect of the density-dependent potential in determining
the limiting potential at lower voltages is striking. In its
absence, the current density tunneling through the field-enhanced
barrier is for example $\ovJ = 3.48 \times 10^{-4}$ at $24.4$V 
corresponding to $\beta \simeq 7$ in the FN formula \cite{details}. 
With the inclusion of the density dependent potential, the limiting
current density drops to $\ovJ \simeq 5.19 \times 10^{-7}$
corresponding to a $\beta \simeq 4.7$ in the FN formula.

The role of the space-charge potential in this lowering is
insignificant at lower voltages since on removing the space-charge 
contribution, the limiting current density does not increase significantly.
Thus, in nanogaps, the exchange-correlation potential plays a significant
role and is likely responsible for the high turn-on voltages reported in
literature  for nanogaps as compared to microgaps \cite{Cheng:nano_res_let}. 

\section{Conclusions}

In conclusion, we have provided a multi-energy-group formalism for 
calculating the field emission current in the presence of electron-electron interaction
based on the existence of a limiting potential.
We have found that at low voltages, the limiting emission current displays FN behaviour.
The study also establishes that the exchange-correlation
potential plays a significant role in lowering the emission current in nanogaps.

\section{References} 



\begin{thebibliography}{99}
\bibitem{contrast} At larger gaps however, the current may be
emission limited at lower voltages since the exchange-correlation contribution is weak.
\bibitem{Ninfinite} Ideally $N \rightarrow \infty$; 
It is chosen to be large and convergence checked.
\bibitem{x0} Bounds on $x_0$ can be estimated based on the value of the 
metal exchange-correlation function in its bulk. Continuity at 
$x=0$ with the bulk value puts a lower bound with $x_0$ determined by
$V_{im} + \phi + E_F = 0$. A loose upper bound can be obtained
from $V_{im} + \phi = 0$. The value of $x_0$ for the field emission
problem probably lies somewhere in between depending on the
value of $r_s$ at $x=0$. We shall use the value of $x_0$ from
$V_{im} + \phi = 0$. This under-estimates the field emission current.
For a discussion, see for instance \cite{kiejna}. The exact value of
$x_0$ is unimportant so far as the formalism presented here is concerned.
\bibitem{neglected_anode} The response of the anode is not
modelled in this calculation since the gap $D = 30$nm.
\bibitem{model_all} Outside the unperturbed metal-vacuum interface, the
asymptotic form of the exchange-correlation potential is image-potential like and 
the image charge is ascribed to the exchange-correlation hole that stays behind 
for an electron in the vacuum region \cite{kiejna}.
\bibitem{negligible} The contribution of the electrostatic potential $\ovV$ in $\ovV_{den}$ is negligible
at lower voltages.
\bibitem{choice} The choice of the enhanced potential is largely unimportant
so far as the multi-group formalism is concerned.
\bibitem{details} The parameters in the floating-sphere model 
result in enhancement of the surface applied
electric field (i.e at $\ovx = 0$) by $53.5$. This 
value decreases rapidly away form the surface. Note that for the 
range of energies at which emission takes place, the slope of the
effective potential is smaller than 1 (see Fig.~1(b)). 
The transmission coefficient nevertheless increases since the barrier
height away from the surface becomes smaller on including enhancement. 
Thus, $\beta$ in the FN-formula exceeds 1.
\end{thebibliography}
\end{document}